\documentclass[a4paper,10pt]{article}
\usepackage[utf8x]{inputenc}
\begin{document}

\Large\begin{center}{\Large \bf   
Where the Lorentz-Abraham-Dirac equation for the radiation reaction force
fails, and why the ``proofs'' break down}
 \\[5 ex]

{\bf Dieter Gromes}\\[3 ex]Institut f\"ur
Theoretische Physik der Universit\"at Heidelberg\\ Philosophenweg 16,
D-69120 Heidelberg \\ E - mail: d.gromes@thphys.uni-heidelberg.de \\
 {\em July 2015}
\end{center} 
\vspace{0.4cm}

{\bf Abstract:} 

We calculate the energy radiated coherently by a system of
$N$ charged non relativistic particles.
It disagrees with the energy loss which is obtained if one employs the
Lorentz Abraham Dirac (LAD) equation for each particle, 
and sums up the contributions. This fact was already clearly stated in 
the classical literature long ago. The reason for the discrepancy is the 
omission of the mixing 
terms in the Poynting vector. For some simple systems we 
present a generalized equation for the 
radiation reaction force which cures this defect.
The counter examples show that the LAD equation 
cannot be generally valid
and that all ``proofs'' must fail somewhere. We demonstrate 
this failure for some
popular examples in the literature.

\section{Introduction}
It is known for more than hundred years now, that an accelerated charge radiates,
and that the energy for this radiation should be provided by a 
radiation reaction force $f^\mu$ which acts like a friction to the 
motion of the particle. The earliest and most popular form of this reaction
 force is the Lorentz-Abraham-Dirac 
(LAD) force \cite{Lorentz}\cite{Abraham}\cite{Dirac},

\begin{equation}
f^\mu =  \frac{2}{3} e²
\big(\ddot{u}^\mu + \dot{u}² u^\mu\big),
\end{equation}
with $u^\mu$ the four velocity, and the dot denoting differentiation with respect 
to the proper time $\tau$.
Due to strange properties like the possible appearance of runaway solutions
one often modifies this equation by introducing the zero order solution of
the equation of motion into the rhs, arriving at the equation, usually called 
the Landau-Lifschitz (LL) equation \cite{LL}:

\begin{equation}
f^\mu =  \frac{2}{3} \frac{e²}{m}
\big(\frac{d}{d\tau}F^\mu +  (F \dot{u})u^\mu\big),
\end{equation}
with $F^\mu$ the external force. See also \cite{Eliezer}\cite{Ford}\cite{Rohrlich}. 
Some of these authors consider this equation as the fundamental
one and even claim it as being ``exact'' or providing the ``correct'' equation.

The non relativistic limit of  equations (1), (2) is
\begin{equation}
 {\bf f} = \frac{2}{3} e² \ddot{\bf v} \approx \frac{2}{3} \frac{e²}{m} \frac{d}{dt} {\bf F}.
\end{equation}
There is still a lively discussion about these equations. In particular there are various
papers which claim to have provided a general proof. This is astonishing, because already 
Abraham \cite{Abraham}, (\& 15, p. 119) clearly states that the equation cannot
be always correct! 
As an example he mentions a number of equally distributed electrons
which move along a circle with equal velocities. Each electron radiates less than
predicted by (1). In the limit of many electrons the motion of the electrons corresponds 
to a stationary current; there is no radiation at all.

The effect has nothing to do with the subtle
differences between equations (1) and (2). The problem arises already for weak fields
where these equations are equivalent, and where the non relativistic limit (3) is
applicable. No complicated mathematics or subtle physics is involved
in the arguments, just old fashioned textbook electrodynamics.

The relevant point is simple. The energy density as well as the Poynting vector
are quadratic in the fields. Therefore it does not make sense to add up the 
energy of the radiation fields created by various particles. One has first to
superimpose the fields, and subsequently calculate energy densities and momentum
flows. This introduces non diagonal mixing terms which can change the standard
arguments considerably. This simple fact, already emphasized
by Abraham, is well known and discussed in books
on accelerator physics \cite{Buts}. But it is usually not even mentioned at all in reviews 
on radiation reaction \cite{Eliezerrev}\cite{Erber}\cite{Rohrrev}\cite{Hammond}. One can 
only find a brief remark in a different
context in \cite{Erber}, p. 382. To our knowledge the problematics  appears to
be completely unknown, forgotten, or ignored by the authors who try to
proof the  LAD- or the LL equation, apparently being unaware that they try to
proof an equation which fails already for simple counter examples. All 
general ``proofs'' must be incorrect.

In the present paper we point out all this in some detail.
In sect. 2 we calculate the radiation of $N$ non relativistic particles and 
show that the application of the LAD equation for each of the particles
would violate energy conservation. In sect. 3, restricting to simple situations,
 we present a
modified equation for the reaction force ${\bf f}_i$ which acts on particle $i$. 
It gives the correct energy balance. Our equation  necessarily depends not 
only on the motion of the particle $i$ under 
consideration, but also on the motion of all the other particles.  
 In sect. 4 we present some simple examples.
Sect. 5 analyzes some of the ``proofs'' in the literature
and shows where they break down. In sect. 6 we investigate under which 
circumstances the standard LAD equation could be correct. 
Sect. 7 gives a summary.

\section{Coherent radiation of $N$ particles}
Consider $N$ particles with masses $m_i$ and charges $e_i$, moving with non relativistic
velocities $v_i \ll c$.   We assume that for early times $t<t_a$
 and late times $t>t_b$ the particles are far apart, such that any interaction and 
acceleration is negligible.  During the intermediate time 
between $t_a$ and $t_b$ the charges can approach each other, 
are accelerated, and can radiate. For large negative times $t<t_a$ there shall be no fields except the 
fields of the freely moving charges. This implies that one
 has to use retarded fields in the following. The  energy which is radiated during the
interval $t_a< t<t_b$ can be calculated by 
slightly generalizing the usual procedure. Choose a  sphere with a large radius $r$, 
 with the particles roughly in the center of the sphere. The non 
relativistic limit of the retarded 
radiation fields created by particle $i$ is given by the well known expressions

\begin{equation}  
 {\bf E}_i({\bf r},t) = e_i \frac{{\bf R}_i \times [{\bf R}_i \times {\bf a}_i(t'_i)]}{R_i³},\quad 
 {\bf B}_i({\bf r},t) = \frac{{\bf R}_i}{R_i} \times  {\bf E}_i({\bf r},t).
\end{equation}
 Here ${\bf R}_i = {\bf r} - {\bf x}_i(t'_i)$ is the difference between the point ${\bf r}$ 
on the sphere where the fields are considered, and the position ${\bf x}_i(t'_i)$   of particle $i$ at
the retarded time $t'_i$, defined implicitly by the relation $R_i = t - t'_i$. The acceleration
${\bf a}_i(t'_i) = d{\bf v}_i(t'_i)/dt'_i$ of particle $i$
enters at the retarded time $t'_i$. If $r>t_b-t_a$, the radiation arrives at the surface of the 
sphere at a late time $>t_b$ where the charges are essentially free and do no longer emit radiation.
Furthermore choose $r$ large compared to the maximal distance $\Delta_{max}$ of 
$ |{\bf x}_i(t'_i) - {\bf x}_j(t'_j)|$ in the interval
$t_a<t'_i,t'_j<t_b$.  Then one has ${\bf R}_i = {\bf r} + O(\Delta_{max})$ for all $i=1,\cdots ,N$. 
One may replace ${\bf R}_i$ by ${\bf r}$, and the relative error can be made as small as one likes, if $r$  is
taken large enough. The total 
radiation fields ${\bf E}({\bf r},t) = \sum_{i=1}^N {\bf E}_i({\bf r},t)$ and
${\bf B}({\bf r},t) = \sum_{i=1}^N {\bf B}_i({\bf r},t)$  become

\begin{equation}
 {\bf E}({\bf r},t) =  \frac{{\bf r} \times [{\bf r} \times \sum _{i=1}^N e_i {\bf a}_i(t'_i)]}{r³},
\quad 
 {\bf B}({\bf r},t) = \frac{{\bf r}}{r} \times  {\bf E}({\bf r},t).
\end{equation}
The retarded times $t'_i$ in (5) are all different, but for non relativistic motions the differences 
of the accelerations at different times are of order $v$. Therefore one can
 replace all retarded times $t'_i$ by some common average retarded time $t'$. 
Indeed this argument needs a more careful investigation which we postpone to sect. 4. 
But for the cases where the argumentation is valid one may replace $t'_i \rightarrow t'$.
Eq. (5) then has the same form as the radiation field of a single particle, with the replacement
 $e {\bf a}(t') \rightarrow \sum _{i=1}^N e_i {\bf a}_i(t')$. Therefore one can simply repeat
the standard derivation of the  Larmor formula and ends up with the following expression for
the energy which was radiated off during the time interval $t_a< t< t_b$:  

\begin{equation}
 E_{rad} = \int_{t_a}^{t_b} \frac{2}{3} (\sum _{i=1}^N e_i {\bf a}_i(t))² dt.
\end{equation}
One could have guessed this formula immediately, but we took care to
derive it explicitly. Indeed such a straightforward generalization 
is only possible for  simple systems. For more complicated N-particle systems as well
as for relativistic motions the 
differences between the accelerations at different times $t'_i$ become
relevant. In this case the derivation of a closed expression like (6) appears hopeless.

\section{Radiation reaction force}
Recall the procedure in the case of a single particle. 
The integral over the work performed by the radiation reaction force ${\bf f}$ 
on the particle is 
identified with the radiated energy, subsequently one performs a partial integration:

\begin{eqnarray}
\lefteqn{\int_{t_a}^{t_b} {\bf v}(t) {\bf f}(t) dt =} \nonumber\\ 
& & - \int_{t_a}^{t_b} \frac{2}{3} e² {\bf a}^2(t) dt  = 
 \int_{t_a}^{t_b} \frac{2}{3} e² {\bf v}(t) \dot{\bf a}(t) dt 
- \frac{2}{3} e² {\bf v}(t) {\bf a}(t) |_{t_a}^{t_b}.
\end{eqnarray}
The last term vanishes if the acceleration vanishes at the endpoints.
By identifying the integrands one obtains the well known non relativistic version 
(3) of the LAD equation, $ {\bf f}(t) =  \frac{2}{3}e²\dot{\bf a}(t)$.
Clearly this formula fails in our case. If one simply would put  
${\bf f}_i(t) =  \frac{2}{3}e_i^2\dot{\bf a}(t)$ for the radiation reaction force acting on 
particle $i$, one would only  
reproduce the diagonal terms $\sim e_i^2 {\bf a}_i^2$ in the energy expression (6), 
but miss all the non diagonal terms
$\sim e_i e_j {\bf a}_i {\bf a}_j$  for $i \ne j$.
This shows already clearly that the LAD equation cannot be always valid.

To obtain a consistent expression for ${\bf f}_i$,  
 we can proceed analogous to the one particle case. There is, however, more freedom 
now, because the mixed terms $\sim e_ie_j$ in (6) can
be compensated by a reaction force acting either on particle $i$ or $j$, or
some combination. 

Due to the symmetry in $i$ 
and $j$ one can write (6) in the form 
\begin{equation}
 E_{rad} = 
\int_{t_a}^{t_b} \frac{2}{3} \sum _{i,j=1}^N 
e_i e_j\lambda_{ij}{\bf a}_i(t) {\bf a}_j(t) dt,
\end{equation}
with $\lambda_{ij}= const$ for simplicity, satisfying 
 $\lambda_{ij} + \lambda_{ji}=2$. In particular 
this implies, of course, $\lambda_{ii}$ ({\em no sum}) = 1. 
The energy expression (8) is obviously independent of the choice of the $\lambda_{ij}$,
but the manifest symmetry in $i,j$ will be broken, after we perform the partial
integration. 

Consider the energy equation analogous to (7), and
perform the partial integration by integrating ${\bf a}_i(t)$
and differentiating ${\bf a}_j(t)$:

\begin{eqnarray}
\int_{t_a}^{t_b} \sum _{i=1}^N {\bf v}_i(t) {\bf f}_i(t) dt =  
- \int_{t_a}^{t_b} \frac{2}{3} \sum _{i,j=1}^N 
e_ie_j \lambda_{ij}{\bf a}_i(t) {\bf a}_j(t)dt \nonumber\\ = 
 \int_{t_a}^{t_b} \frac{2}{3} \sum _{i,j=1}^N  e_ie_j
 \lambda_{ij}{\bf v}_i(t) \dot{\bf a}_j(t) dt 
- \frac{2}{3} \sum _{i,j=1}^N  e_i e_j
 \lambda_{ij}{\bf v}_i(t) {\bf a}_j(t) |_{t_a}^{t_b}.
\end{eqnarray}
Again the last term can be made arbitrarily small if the times 
$t_a,t_b$ are chosen appropriately, and a possible solution for the reaction forces 
which fulfills (9) is

\begin{eqnarray}
 & & {\bf f}_i(t) =   \frac{2}{3} e_i \sum _{j=1}^N e_j\lambda_{ij}
\dot{\bf a}_j(t),\\
& & \mbox{with \quad} \lambda_{ij} + \lambda_{ji}(t)=2. \nonumber
\end{eqnarray}
The term in the sum with $j=i$ is identical with the usual non relativistic form of
the LAD force. But there are necessarily also the additional terms which involve 
the accelerations of the other particles. 

We mention that Eliezer \cite{Eliezerrev}, (30) - (32), gives an equation 
of motion for a system of many electrons. It is not
identical with that obtained from our eq. (10), because it only 
contains the Lorentz forces arising from the other charges, but not the 
$\dot{\bf a}_j(t)$ for $j\neq i$, thus violating energy conservation.

\section{Examples}
Consider first a system of two particles. Here one can rigorously justify
the replacements $t'_i \rightarrow t'$ in (5)  without any further assumptions. 
The difference of the 
retarded times is at most as large as the distance $x$ between the charges, 
$|t'_1-t'_2| \leq  x$. It is irrelevant whether one uses the distance at equal 
times or at the respective retarded times, they differ at most by a factor (1+$v$).
 In leading order the force is the Coulomb force. 
Shorthand, suppressing irrelevant vector notation or factors of order 1,
 one has for any of the accelerations $ma \sim e^2/ x^2,\quad
m\dot{a} \sim(e²/x^3)v$ i.e. $\dot{a}/a \sim v/x$. From the mean value theorem one 
then gets $\ln (a(t'_i)/ a(t')) = \ln (a(t'_i)/a_0) - \ln (a(t')/a_0) =
(t'_i-t')\dot{a}(\bar{t})/a(\bar{t}) \sim  v$, with $\bar{t}$ some time between
$t'_i$ and $t'$. Therefore $a(t'_i)/a(t')\sim(1+v)$, and it is legitimate to 
 replace ${\bf a}_i(t'_i)$ by  ${\bf a}_i(t')$.

 If we specialize to particles with identical masses, 
$m_1=m_2$, and with charges either
identical or opposite, $e_1 = \pm e_2$, symmetry considerations can fix
the $\lambda_{ij}$. Symmetry between the particles in 
the first case, and charge conjugation symmetry in the second case, now 
 implies $\lambda_{ij}=1$. 
 The center of mass moves with constant velocity, the sum of the 
accelerations vanishes,  $\sum_i \bf{a}_i = 0$.  

For the case of identical charges, eq. (6) then implies that there is no
radiation at all. Consequently there is also no radiation reaction force, 
in spite of the fact that the particles are accelerated!

Opposite charges provide an example of
the other extreme. Each particle experiences a reaction force 
which is twice the LAD force (3).

The two particle case is the only one where one can immediately
justify the performed approximations without  further assumptions.
The reason is that the time difference, the distance of
the particles, and the acceleration are directly connected. If the 
acceleration and its derivative  is large, the time difference 
is small. Vice versa, if the time difference is large, the acceleration is
small. 

For a multi particle system this is not necessarily the case. 
Consider, e.g. a four particle system where (1,2) as well as (3,4) are close 
together, while (1,2) are far apart from (3,4). All accelerations and their
derivatives, but also the time differences $t'_{1,2} - t'_{3,4}$ are
large. The relative error from the replacements 
${\bf a}_i(t'_i) \rightarrow {\bf a}(t')$ is still of 
order $v$, but the factor in front may become inadmissible large.
It needs a special investigation whether one can do the replacement.
Essentially the approximation should be valid if all particles have similar distances.

The expressions (5) for the fields stay valid for all (non relativistic) 
multi particle systems, but already the calculation of the radiated energy would
become clumsy if one keeps the different times $t'_i$. It appears unlikely that
a closed formula for the reaction forces can be derived. 

In all cases one has to be aware that there are always mixing terms,
even between charges which are rather far apart. One should understand
how the mixing terms decrease if the separation increases. Consider
two particles $i,j$ which both essentially radiate only between $t_a$ and $t_b$.
The distance $x_{ij}$ may be very large, even larger than $t_b-t_a$. On 
most points of the sphere with the large radius $r$ there is no interference,
because the radiation from the two charges cannot arrive at the same time. Only
in directions which are essentially orthogonal to ${\bf x}_{ij}$ one has
$t'_i \sim t'_j$, there will be always interference. If $x_{ij}$
increases  these regions become smaller, so finally the fields from the two
 charges decouple. 

A  remark concerning causality appears appropriate. The mutual influence
of the various charges  does not
violate causality. It even would not violate it in a relativistic
treatment. The use of retarded potentials poses initial conditions at
$t=-\infty$, therefore the whole time evolution is fixed, in particular
also all the accelerations ${\bf a}_j(t)$ are completely determined from
 the initial conditions. No chance to send signals from $j$ to $i$.

\section{Where ``proofs'' fail}
The simple counter example already mentioned by Abraham, together with the examples 
just presented more explicitly in sect. 4,  
clearly and irrefutably demonstrate that the LAD form of the radiation reaction force
cannot be universally valid. A strong but unavoidable consequence is that
all general ``proofs'' of this formula must be incorrect.
In view of this it appears appropriate to point out where proofs fail.
It is, of course, impossible to cover the vast literature here, 
and it is as well impossible to analyze proofs which
are unintelligible, at least for the present author. We concentrate on some 
classical literature and on some more recent papers which claim a proof for the LAD force.
 We emphasize that it is not our intention to criticize the authors of these 
often rather ingenious considerations. But given the
undeniable fact that there are situations where the LAD equation definitely fails,
it appears necessary to show why these considerations cannot be generally correct.
We will sometimes modify the original notation to our conventions.

The first type of approaches identifies the work done by the radiation 
reaction force with the energy radiated off from the particle by the Larmor formula.
From this one extracts the reaction force. There are two problems in this approach.
The counter examples presented above show the first of these problems.
Neither the radiated energy nor the reaction force can be treated separately for
each particle. They are collective phenomena and  necessarily involve
other particles as well. 

The second problem is that energy can be stored and 
released in the electromagnetic field. Although we know, of course, 
the expressions for energy and momentum density of an electromagnetic
field, this does not help to calculate the total energy, because there
are the non integrable singularities at the position of the point particle.
Unless one is willing to introduce extended particles, an alternative which we 
discuss below, apparently the only way to avoid this problem is to restrict the discussion
of energy to  times where the particle is far away from the ``external'' field,
or from the fields of other particles. In these regions the particle moves with constant 
velocity, and the divergent field of
the particle poses no problem. The whole four momentum of the system 
consisting of particle plus its own field is simply given by $P^\mu =m u^\mu$.
Thus one can only derive necessary conditions which make statements on the energy balance 
between times $t_a$ and $t_b$ where the particle under consideration is outside of
all other fields. Therefore, notwithstanding the previous difficulty, one only obtains
an integrated relation. One cannot 
derive relations for a definite time $t$ where the particle
moves within the field. We therefore also cannot maintain all the statements in
our previous paper \cite{GromesThommes}. But we definitely reemphasize the point
made there, that the reaction force depends not only on the motion of
the particle of interest, but also on the way how the acceleration is achieved.

A more direct approach tries to derive the equation of motion for an electron by 
taking into account both an external as well as its own radiation field.
This procedure is again plagued by the singularity of the
 electromagnetic field at the position of the particle. There are two
strategies in order to attack this problematics.

The first method, first employed by Lorentz \cite{Lorentz} and Abraham \cite{Abraham}
and still advocated for by some modern authors \cite{structureFOC}\cite{structureRohr}, 
is to treat the electron as
an extended particle. The original idea of Lorentz, that the mass of the 
electron should be of purely electromagnetic origin, was already criticized by
Abraham. One needs in addition some other forces in order
to hold the whole structure together. 
Since a rigid charge distribution is incompatible with special relativity,
the determination of this distribution would become part of the dynamical problem
and depend on all the forces present in the system. We don't know of any detailed 
model of this kind.
The problems connected with an extended particle appear to be at least as serious
as those arising from the singular field around a point particle.

In a more recent paper Gralla, Harte, and Wald \cite{GHW} claim
``A rigorous derivation ...''. They argue 
that a point like limit of a particle with fixed mass and charge is physically 
impossible. Therefore they consider an extended particle where in the point like
limit mass and charge go to zero, hardly a candidate for a physical electron.
We are not able to comment on their arguments.

If one wants to avoid a model with an extended particle one needs some other limiting procedure. 
Let us take a look on the classical derivation of Dirac \cite{Dirac}.
He uses the singularity free and time symmetric combination
$f^{\mu\nu} = (F^{\mu\nu}_{in} + F^{\mu\nu}_{out})/2$, but 
this is not the relevant point here. Dirac derives the equation 
$\frac{1}{2} e^2\epsilon^{-1}\dot{u}^\mu - e u_\nu f^{\mu\nu}
=\dot{B}^\mu$,
where $\epsilon$ is the small radius of a tube surrounding the electron.
Obviously one must have $u \dot{B} =0$. This $B^\mu$ cannot be 
calculated, therefore Dirac makes use of the most simple ansatz 
$B^\mu = k u^\mu$, discarding more complicated expressions like
e.g. $B^\mu = k' [\dot{u}^4 u^\mu + 4 (\dot{u} \ddot{u}) \dot{u}^\mu]$.
In fact, there are more general possibilities as given by 
Eliezer \cite{Eliezerrev}. The result for the force is ambiguous.
 
An interesting procedure was suggested by Barut \cite{Barut}.
To avoid the divergences of the electromagnetic field at the position
 $x^\mu (t)$
of the particle he starts with the Lorentz force at a slightly displaced position,
with the field taken at 
$y^\mu= x^\mu (t+\epsilon)=x^\mu (t) + \epsilon \dot{x}^\mu (t) 
+ \epsilon² \ddot{x}^\mu (t) /2 + \epsilon ³ \stackrel{...}{x}^\mu (t)/6 + \cdots$.
A possible singular term $\sim 1/\epsilon²$ in the force 
vanishes, the next singular term $ \sim 1/\epsilon$ is absorbed into
a mass renormalization, while the term $ \sim \epsilon⁰$ gives the LAD reaction force.
All the higher terms vanish in the limit $\epsilon \rightarrow 0$.

This approach appears elegant and convincing, but one has to realize that it uses
quite a special choice for the limiting procedure. Even if one is willing to
approach the point $x^\mu$ along the trajectory, one could use a more general
limiting procedure of the form $y^\mu =  x^\mu + \epsilon x^\mu_{(1)}
 +  \epsilon² x^\mu_{(2)}/2
+ \epsilon^3 x^\mu_{(3)}/6 + \cdots$. 
In order to remove the singularity $\sim 1/\epsilon²$ one has to put
$x^\mu_{(1)} = \dot{x}^\mu (t)$ as before. The next order $\sim 1/\epsilon$ is again
 absorbed by mass renormalization. But one is completely free to choose
the function $x^\mu_{(3)}$ as one likes. Instead of (1) one then ends up with 
$f^\mu =  \frac{2}{3} e²\big(x_{(3)}^\mu -  (u x_{(3)}) u^\mu \big)$. 

In general one may say that all ``derivations'' use some special ansatz 
and tacitly assume that the
 reaction force depends on the motion of the considered particle only,
and on nothing else. We have seen that this is in general not true.

\section{When can the LAD force be correct?}
A hasty and naive comment on the counter examples given above could consist
in postulating that the various derivations in the literature should (of course!) 
be 
only applicable in the case of an external field. Whatever the arguments
for such a restriction could be, it would require a clear definition of
an external field. This is more tricky than one might expect.

A first try for a definition could be: An external field is a field which 
is not changed by the presence of the test particle. This would be the 
case in the limit that the charge $e$ of the test particle
goes to zero, keeping everything else fixed. Our examples clearly
show that this would not help. To the contrary, the mixing terms $\sim ee_j$
would now become dominant compared to the diagonal term $\sim e^2$.

The examples also show that it would not help to define an external field
as being a field which is created by many other particles. 

A definition, though somewhat academic, which might work, is to create the field
by one or more charged particles with a very large mass compared to 
the mass of the test particle.  Of course, no further light particles should
be around. To make this more explicit consider a two particle system 
 with $m_1 \ll m_2$. Using $m_1 \dot{\bf a}_1 + m_2 \dot{\bf a}_2=0$
 one can write the reaction forces (10) as

\begin{eqnarray}
 {\bf f}_1 & = &  \frac{2}{3} (e_1^2 -\lambda_{12}\frac{m_1}{m_2}e_1e_2 )
\dot{\bf a}_1, \nonumber\\
{\bf f}_2 & = &  \frac{2}{3} (e_2^2 -\lambda_{21}\frac{m_2}{m_1}e_1e_2 )
\dot{\bf a}_2.
\end{eqnarray}
If one had $\lambda_{ij} = 1$, the formula for the light particle would
approach the LAD form, while that for the heavy particle would contain a large
factor compared to LAD. If, on the other hand, one would have, e.g.
$\lambda_{ij}= 2 m_j^2/(m_i^2 + m_j^2)$, both forces would approach the LAD 
form in the limit $m_1/m_2 \rightarrow 0$.

Whenever one talks about an external
field (which breaks, by the way, momentum conservation, and, if time dependent,
also energy conservation) one should specify how this field is realized and
analyze the whole system of particle plus ``external'' field.

A realistic chance of really measuring the radiation reaction 
force can probably only come from the use of intense laser beams. There has been some
discussion on this recently \cite{Laser}. At present we are 
unable to discuss the consequences of our considerations for this case. The
motion of the electron is necessarily relativistic, besides the electron of
interest there is a huge number of other electrons, part of which producing
the coherent laser radiation, a process which apparently requires the 
consideration of quantum mechanics. The situation is complex, we 
don't try to formulate a guess here.

\section{Summary and conclusions}

We called attention to some facts which are almost trivial,
but have, to our knowledge, never been carefully
exploited with respect to their consequences. Radiation reaction cannot
be treated as a phenomenon which only concerns a particle and an external field.
Energy density and Poynting vector are quadratic
in the fields, therefore they are not simply obtained by summing up the 
contributions of all particles involved. There are mixing terms. 
This implies that the energy loss through 
the friction of the Lorentz Abraham Dirac force is usually not identical 
with the radiated energy. All general ``proofs`` of the LAD force are thus
bound to fail. In (10) we suggested a form of the reaction force which 
respects energy conservation, but such a closed expression  
can be derived only for simple
non relativistic systems. The question how to treat realistic situations
like an electron in a strong laser field remains  open.

\vspace{1cm} 
{\bf Acknowledgement: } I thank E. Thommes for valuable discussions 
and a careful reading of the manuscript.

\end{document}